\documentclass[prl,aps,reprint]{revtex4-1}

\usepackage{amsmath}
\usepackage{amssymb}
\usepackage{graphicx}
\usepackage{pict2e}
\usepackage{color}
\usepackage{hyperref}

\newcommand{\sgn}{\operatorname{sgn}}
\newcommand{\C}[2]{{#1 \choose #2}}
\newcommand{\Li}{\operatorname{Li}}

\renewcommand{\Re}{\operatorname{Re}}
\renewcommand{\Im}{\operatorname{Im}}

\newcommand{\rme}{\mathrm{e}}
\newcommand{\rmi}{\mathrm{i}}
\newcommand{\rmd}{\mathrm{d}}

\begin{document}

\title{Finite-time fluctuations for the totally asymmetric exclusion process}
\author{Sylvain Prolhac}
\email{sylvain.prolhac@irsamc.ups-tlse.fr}
\affiliation{Laboratoire de Physique Th\'eorique, Universit\'e de Toulouse, UPS, CNRS, France}

\begin{abstract}
The one-dimensional totally asymmetric simple exclusion process (TASEP), a Markov process describing classical hard-core particles hopping in the same direction, is considered on a periodic lattice of $L$ sites. The relaxation to the non-equilibrium steady state, which occurs on the time scale $t\sim L^{3/2}$ for large $L$, is studied for the half-filled system with $N=L/2$ particles. Using large $L$ asymptotics of Bethe ansatz formulas for the eigenstates, exact expressions depending explicitly on the rescaled time $t/L^{3/2}$ are obtained for the average and two-point function of the local density, and for the current fluctuations for simple (stationary, flat and step) initial conditions, relating previous results for the infinite system to stationary large deviations. The final formulas have a nice interpretation in terms of a functional integral with the action of a scalar field in a linear potential.
\end{abstract}


\maketitle

Non-equilibrium statistical mechanics aims to explain the emergence at large scales of simple evolution laws for a few macroscopic variables from the microscopic dynamics of many degrees of freedom. Universality appears at the mesoscopic scale, when the system is large but still finite, as only essential features of microscopic randomness subsist in the fluctuations of the macroscopic variables.

In specific classes of systems with local microscopic dynamics, fluctuations of macroscopic observables such as the position of the interface in growth models, the current of particles in driven lattice gases or the free energy for directed polymers in random media, are described by KPZ \cite{KPZ1986.1} universality \cite{KK2010.1,QS2015.1,HHT2015.1}. The amplitude of KPZ fluctuations grows as $t^{\alpha/z}$ in the early time regime and saturates at $\ell^{\alpha}$ in a region of space of size $\ell$, with a crossover on the time scale $t\sim\ell^{z}$. The roughness and dynamical exponents, $\alpha=1/2$ and $z=3/2$ in one dimension, were measured experimentally in growing bacterial colonies \cite{MWIRMSM1998.1} and burning fronts \cite{MMAAMT2001.1}. Scaling functions obtained during the last 15 years from several exactly solvable models were also observed in turbulent liquid crystal experiments \cite{TS2010.1,TSSS2011.1} and in numerical studies of various models \cite{LK2006.1,T2012.1,HHL2014.1}.

The one-dimensional totally asymmetric simple exclusion process (TASEP) \cite{D1998.1,GM2006.1} is an interacting particle system that has been used as a basis for models of cellular molecular motors \cite{CMZ2011.1}, traffic flow \cite{CSS2000.1}, chains of quantum dots \cite{KvO2010.1}, and whose fluctuations belong to KPZ universality. TASEP describes the movement of hard-core particles, with the exclusion constraint that each site is either occupied by a single particle or empty. Particles move with a continuous-time Markovian dynamics, hopping from any site $i$ to the next site $i+1$ with rate $1$ if the destination site is empty. We consider in this letter TASEP with $N$ particles on a periodic lattice of $L$ sites, on the crossover time scale $t\sim L^{z}$ which describes the evolution of KPZ fluctuations from those of the infinite system to the ones in the non-equilibrium steady state.

TASEP is a stochastic integrable model \cite{S2012.1} with generator equal to the non-Hermitian Hamiltonian of a twisted XXZ spin chain. The formal analogy with quantum integrability provides powerful Bethe ansatz tools to analyse the model. Furthermore, TASEP always relaxes to its unique stationary state, avoiding the complicated issues about thermalization occurring in quenches of quantum integrable systems with unitary evolution.

The probability $P_{t}(\mathcal{C})$ to observe at time $t$ a configuration $\mathcal{C}$ of TASEP obeys a master equation, and $P_{t}(\mathcal{C})=\langle\mathcal{C}|\rme^{tM}|\mathcal{C}_{0}\rangle$ with $M$ the Markov matrix, for a system starting in configuration $\mathcal{C}_{0}$ at time $0$. We focus on three types of analytically tractable initial conditions: the stationary state where all configurations are equally likely, a flat configuration with equally spaced particles, and a step configuration with $N$ consecutive sites occupied. The average of the number of particles $\eta_{i}(\mathcal{C})$ (equal to $0$ or $1$) at site $i$ is given by
\begin{equation}
\label{Average eta}
\langle\eta_{i,t}\rangle=\sum_{\mathcal{C}}\openone_{\{\eta_{i}(\mathcal{C})=1\}}\langle\mathcal{C}|\rme^{tM}|\mathcal{C}_{0}\rangle\;,
\end{equation}
and similarly for the two-point function $\langle\eta_{0,0}\eta_{i,t}\rangle$.

\begin{figure}
  \begin{center}
    \begin{tabular}{ccccc}
      \begin{picture}(28,10)(-3,0)
        \put(20.5,-1){$x$}
        \put(1.2,8.5){$\rho(x,0)$}
        \put(0,0){\vector(1,0){20.5}}
        \put(0,0){\vector(0,1){10}}
        \put(0.,2.6){\line(0.2,0.8){0.2}}\put(0.2,3.4){\line(0.2,0.8){0.2}}\put(0.4,4.2){\line(0.2,-1.6){0.2}}\put(0.6,2.6){\line(0.2,2.4){0.2}}\put(0.8,5.){\line(0.2,0.6){0.2}}\put(1.,5.6){\line(0.2,-2.2){0.2}}\put(1.2,3.4){\line(0.2,1.){0.2}}\put(1.4,4.4){\line(0.2,0.2){0.2}}\put(1.6,4.6){\line(0.2,1.2){0.2}}\put(1.8,5.8){\line(0.2,-0.2){0.2}}\put(2.,5.6){\line(0.2,-0.2){0.2}}\put(2.2,5.4){\line(0.2,-1.6){0.2}}\put(2.4,3.8){\line(0.2,0.2){0.2}}\put(2.6,4.){\line(0.2,-1.4){0.2}}\put(2.8,2.6){\line(0.2,3.2){0.2}}\put(3.,5.8){\line(0.2,-2.4){0.2}}\put(3.2,3.4){\line(0.2,-0.2){0.2}}\put(3.4,3.2){\line(0.2,0.4){0.2}}\put(3.6,3.6){\line(0.2,0.){0.2}}\put(3.8,3.6){\line(0.2,-1.6){0.2}}\put(4.,2.){\line(0.2,2.){0.2}}\put(4.2,4.){\line(0.2,-0.2){0.2}}\put(4.4,3.8){\line(0.2,-1.4){0.2}}\put(4.6,2.4){\line(0.2,-0.6){0.2}}\put(4.8,1.8){\line(0.2,1.6){0.2}}\put(5.,3.4){\line(0.2,-1.4){0.2}}\put(5.2,2.){\line(0.2,0.8){0.2}}\put(5.4,2.8){\line(0.2,2.6){0.2}}\put(5.6,5.4){\line(0.2,-2.2){0.2}}\put(5.8,3.2){\line(0.2,0.){0.2}}\put(6.,3.2){\line(0.2,1.4){0.2}}\put(6.2,4.6){\line(0.2,-0.4){0.2}}\put(6.4,4.2){\line(0.2,0.6){0.2}}\put(6.6,4.8){\line(0.2,0.){0.2}}\put(6.8,4.8){\line(0.2,-0.6){0.2}}\put(7.,4.2){\line(0.2,0.4){0.2}}\put(7.2,4.6){\line(0.2,-0.2){0.2}}\put(7.4,4.4){\line(0.2,0.){0.2}}\put(7.6,4.4){\line(0.2,0.2){0.2}}\put(7.8,4.6){\line(0.2,-1.){0.2}}\put(8.,3.6){\line(0.2,-1.){0.2}}\put(8.2,2.6){\line(0.2,0.8){0.2}}\put(8.4,3.4){\line(0.2,0.6){0.2}}\put(8.6,4.){\line(0.2,1.){0.2}}\put(8.8,5.){\line(0.2,-0.4){0.2}}\put(9.,4.6){\line(0.2,0.6){0.2}}\put(9.2,5.2){\line(0.2,-0.2){0.2}}\put(9.4,5.){\line(0.2,-1.8){0.2}}\put(9.6,3.2){\line(0.2,0.6){0.2}}\put(9.8,3.8){\line(0.2,-0.6){0.2}}\put(10.,3.2){\line(0.2,1.6){0.2}}\put(10.2,4.8){\line(0.2,-0.4){0.2}}\put(10.4,4.4){\line(0.2,-1.){0.2}}\put(10.6,3.4){\line(0.2,-0.4){0.2}}\put(10.8,3.){\line(0.2,0.8){0.2}}\put(11.,3.8){\line(0.2,1.){0.2}}\put(11.2,4.8){\line(0.2,-3.){0.2}}\put(11.4,1.8){\line(0.2,3.){0.2}}\put(11.6,4.8){\line(0.2,-0.2){0.2}}\put(11.8,4.6){\line(0.2,-1.8){0.2}}\put(12.,2.8){\line(0.2,0.4){0.2}}\put(12.2,3.2){\line(0.2,-0.2){0.2}}\put(12.4,3.){\line(0.2,-1.2){0.2}}\put(12.6,1.8){\line(0.2,0.6){0.2}}\put(12.8,2.4){\line(0.2,2.4){0.2}}\put(13.,4.8){\line(0.2,1.4){0.2}}\put(13.2,6.2){\line(0.2,-1.){0.2}}\put(13.4,5.2){\line(0.2,-0.4){0.2}}\put(13.6,4.8){\line(0.2,0.2){0.2}}\put(13.8,5.){\line(0.2,0.2){0.2}}\put(14.,5.2){\line(0.2,1.){0.2}}\put(14.2,6.2){\line(0.2,-1.){0.2}}\put(14.4,5.2){\line(0.2,0.6){0.2}}\put(14.6,5.8){\line(0.2,-1.6){0.2}}\put(14.8,4.2){\line(0.2,-0.4){0.2}}\put(15.,3.8){\line(0.2,0.8){0.2}}\put(15.2,4.6){\line(0.2,-0.8){0.2}}\put(15.4,3.8){\line(0.2,-0.4){0.2}}\put(15.6,3.4){\line(0.2,1.2){0.2}}\put(15.8,4.6){\line(0.2,0.){0.2}}\put(16.,4.6){\line(0.2,-2.){0.2}}\put(16.2,2.6){\line(0.2,0.2){0.2}}\put(16.4,2.8){\line(0.2,1.2){0.2}}\put(16.6,4.){\line(0.2,-1.8){0.2}}\put(16.8,2.2){\line(0.2,1.6){0.2}}\put(17.,3.8){\line(0.2,1.4){0.2}}\put(17.2,5.2){\line(0.2,-2.4){0.2}}\put(17.4,2.8){\line(0.2,3.2){0.2}}\put(17.6,6.){\line(0.2,-2.2){0.2}}\put(17.8,3.8){\line(0.2,-1.2){0.2}}
        \put(-2.5,3){\footnotesize$\overline{\rho}$}
        \put(-0.5,4){\line(1,0){1}}
        \put(16.7,-2.5){\footnotesize$1$}
        \put(18,-0.5){\line(0,1){1}}
        \put(19.5,3.7){$\frac{1}{\sqrt{L}}$}
        \put(19,3){\vector(0,1){4}}
        \put(19,6){\vector(0,-1){4}}
      \end{picture}
      &&
      \begin{picture}(25,10)(-3,0)
        \put(20.5,-1){$x$}
        \put(1.2,8.5){$\rho(x,0)$}
        \put(0,0){\vector(1,0){20.5}}
        \put(0,0){\vector(0,1){10}}
        \put(0,4){\line(1,0){18}}
        \put(-2.5,3){\footnotesize$\overline{\rho}$}
        \put(-0.5,4){\line(1,0){1}}
        \put(16.7,-2.5){\footnotesize$1$}
        \put(18,-0.5){\line(0,1){1}}
      \end{picture}
      &&
      \begin{picture}(25,10)(-3,0)
        \put(20.5,-1){$x$}
        \put(1.2,8.5){$\rho(x,0)$}
        \put(0,0){\vector(1,0){20.5}}
        \put(0,0){\vector(0,1){10}}
        \put(10.5,0){\line(0,1){7}}
        \put(10.5,7){\line(1,0){6}}
        \put(16.5,7){\line(0,-1){7}}
        \put(13,9){$\overline{\rho}$}
        \put(11.5,8){\vector(1,0){5}}
        \put(15.5,8){\vector(-1,0){5}}
        \put(-2,6){\footnotesize$1$}
        \put(-0.5,7){\line(1,0){1}}
        \put(16.7,-2.5){\footnotesize$1$}
        \put(18,-0.5){\line(0,1){1}}
      \end{picture}
    \end{tabular}
  \end{center}
  \caption{Stationary, flat and step initial density profiles.}
  \label{Fig init state}
\end{figure}
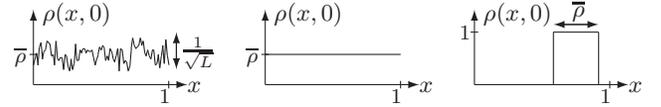

On the hydrodynamic time scale $t\sim L$, the density profile $\rho(x,\tau)$, obtained at large $L$ from the occupation numbers $\eta_{i}$ with $x\simeq i/L$ and $\tau=t/L$, becomes deterministic at leading order: it is solution of Burgers' equation $\partial_{\tau}\rho+\partial_{x}(\rho(1-\rho))=0$. With $\overline{\rho}$ the conserved average density, the initial condition $\rho(x,0)$ is equal in the stationary case to $\overline{\rho}$ plus a Brownian bridge of order $1/\sqrt{L}$, in the flat case to $\overline{\rho}$, and in the step case to $1$ in an interval of length $\overline{\rho}$ and $0$ otherwise, see figure \ref{Fig init state}.

In the first two cases, $\rho(x,\tau)\simeq\overline{\rho}$ at leading order for all $\tau$. In the step case, as with any generic initial condition, Burgers' equation generates shocks (discontinuities) after a finite time $\tau$. Only one shock remains at large $\tau$, moving with velocity $1-2\overline{\rho}$. We consider in the following only the half-filled case $\overline{\rho}=1/2$ to avoid the need of a moving reference frame. Then $\rho(x,\tau)\simeq\frac{1}{2}-\frac{x-\kappa}{2\tau}$ at large $\tau$ for $\kappa-1/2<x<\kappa+1/2$, with $\kappa\pm\frac{1}{2}$ the position of the shock. Our first main result (\ref{sigma average step}), see also figure \ref{Fig aRho cQ}a, describes the average density profile on the time scale $t\sim L^{3/2}$ when the sharp shock vanishes due to fluctuations in the microscopic model.

Multiple point correlations give informations about fluctuations beyond the hydrodynamics. They have been much studied for TASEP on the infinite line $\mathbb{Z}$ \cite{C2011.1}. In particular, the stationary two-point function of the density behaves \cite{PS2002.1} in the long time limit as $\langle\eta_{0,0}\eta_{i,t}\rangle-\frac{1}{4}\simeq2^{-7/3}t^{-2/3}f_{\text{KPZ}}(2^{-1/3}t^{-2/3}i)$. The Pr\"ahofer-Spohn scaling function $f_{\text{KPZ}}$ \footnote{$f_{\text{KPZ}}(y)$ is equal to $g''(y/2)/16$ in the notations of \cite{PS2002.1} and to $g''(y)/4$ in that of $\cite{PrahoferWeb}$} also describes the stationary two-point function for propagating sound modes in generic one-dimensional Hamiltonian systems \cite{vB2012.1}. Our second main result (\ref{sigma 2pt stat}), see also figure \ref{Fig 2ptRho}, is the two-point function for periodic TASEP on the time scale $t\sim L^{3/2}$. It agrees numerically with $f_{\text{KPZ}}$ at short time.

KPZ universality also describes the current of particles of TASEP. The total (time-integrated) current $Q$ between time $0$ and time $t$, equal to $0$ when $t=0$, is incremented by $1$ each time a particle hops anywhere in the system. The generating function of $Q$ verifies a deformed master equation \cite{DL1998.1}. With fugacity $\lambda$ conjugate to $Q$, one has $\langle\rme^{\lambda Q/L}\rangle=\sum_{\mathcal{C}}\langle\mathcal{C}|\rme^{tM(\lambda)}|\mathcal{C}_{0}\rangle$, where $M(\lambda)$ is obtained from multiplying by $\rme^{\lambda/L}$ the non-diagonal entries of $M$ in configuration basis. The local (time-integrated) current $Q_{i}$, which counts only hops between site $i$ and $i+1$, is related to $Q$ by $Q/L=Q_{i}+\frac{1}{L}\sum_{j=1}^{N}([x_{j}]_{i}-[x_{j}^{0}]_{i})$. The $[x_{j}^{0}]_{i}$'s and $[x_{j}]_{i}$'s are integers between $1$ and $L$ counting the positions of the particles from site $i+1$, respectively in the initial configuration $\mathcal{C}_{0}$ and in the final configuration $\mathcal{C}$. Calling $S_{i}$ the diagonal operator on configurations such that $S_{i}|\mathcal{C}\rangle=\sum_{j=1}^{N}([x_{j}]_{i}-L/2)|\mathcal{C}\rangle$, one has
\begin{equation}
\label{GF Qi}
\langle\rme^{\lambda Q_{i}}\rangle=\sum_{\mathcal{C}}\langle\mathcal{C}|\rme^{-\frac{\lambda}{L}\,S_{i}}\rme^{tM(\lambda)}\rme^{\frac{\lambda}{L}\,S_{i}}|\mathcal{C}_{0}\rangle\;.
\end{equation}
The statistics of $Q_{i}$ beyond its stationary value, $Q_{i}\simeq t/4$ for the half-filled system, has been studied extensively for TASEP on the infinite line \cite{C2011.1}, and for a finite system in the long time limit \cite{D2007.1}. Our third main result is the crossover (\ref{xi GF stat})-(\ref{xi GF step}) on the time scale $t\sim L^{3/2}$ between these two regimes, see also figures \ref{Fig aRho cQ}b-\ref{Fig aRho cQ}d and \ref{Fig PDF Qloc stat flat step}b-\ref{Fig PDF Qloc stat flat step}d.

On the infinite line, current fluctuations are defined as $\xi_{t}^{\mathbb{Z}}=(Q_{0}-t/4)/t^{1/3}$, with the amplitude of KPZ fluctuations growing as $t^{1/3}$. For stationary initial state, where sites are occupied independently with probability $1/2$, the statistics of $\xi_{t}^{\mathbb{Z}}$ is given \cite{FS2006.1} in the long time limit by $P(-\xi_{t}^{\mathbb{Z}}<u)\to F_{0}(2^{4/3}u)$ with $F_{0}$ the Baik-Rains distribution \cite{BR2000.1}. For flat initial condition, where every other site is occupied, the statistics is \cite{S2005.1} $P(-\xi_{t}^{\mathbb{Z}}<u)\to F_{1}(4u)$ with $F_{1}$ the Tracy-Widom distribution for the Gaussian orthogonal ensemble (GOE). Finally, for step initial condition, where only sites $i<0$ are occupied, one has \cite{J2000.1} $P(-\xi_{t}^{\mathbb{Z}}<u)\to F_{2}(2^{4/3}u)$ with $F_{2}$ the Tracy-Widom distribution for the Gaussian unitary ensemble (GUE). Both $F_{1}$ and $F_{2}$ appeared initially in random matrix theory, where they describe statistics of the largest eigenvalue. The functions $F_{0}$, $F_{1}$, $F_{2}$ can be expressed in terms of the Hastings-Mcleod solution of Painlev\'e II equation. Alternative expressions in terms of Fredholm determinants with Airy kernels also exist. Numerical evaluations are available at \cite{PrahoferWeb}, see figure \ref{Fig PDF Qloc stat flat step}a.

For a large but finite system, on the other hand, KPZ fluctuations are independent of the initial condition in the long time limit, and their amplitude saturates at order $L^{\alpha}$. The current fluctuations of periodic TASEP are then defined as $\xi_{t}^{\text{st}}=(Q_{i}-t/4)/\sqrt{L}$, which converges almost surely to $t/(4L^{3/2})$ for long $t$. Beyond this deterministic value, $P\big(\xi_{t}^{\text{st}}=\frac{ut}{4L^{3/2}}\big)\sim\exp\big(-tL^{-3/2}\,G_{\text{st}}(u)\big)$ with $G_{\text{st}}$ the Derrida-Lebowitz large deviation function \cite{DL1998.1}, whose Legendre transform has an exact parametric expression. $G_{\text{st}}$ also describes the fluctuations of the free energy for a directed polymer in a random medium \cite{BD2000.1}, of the height for an avalanche model \cite{PPH2003.1}, and of the current for open TASEP at the edge of the maximal current phase \cite{GLMV2012.1}.

Our main results for density and current fluctuations at finite rescaled time $\tau=t/(2L^{3/2})$ follow from expanding (\ref{Average eta}), (\ref{GF Qi}) over the eigenstates $r$ of $M(\lambda)$. Exact Bethe ansatz formulas \cite{B2009.1,MSS2012.2} allow to compute explicitly the large $L$ limit \cite{P2014.1,P2015.2}, see supplemental material. The final expressions (\ref{sigma average step})-(\ref{xi GF step}) have all roughly the form of discretized functional integrals with the action of a scalar field in a linear potential, and various operators inserted depending on the observable and the initial state:
\begin{equation}
\label{functional integral}
\sum_{r}\mathcal{A}_{r}\,\exp\Big(\int_{-\infty}^{\nu_{r}}\rmd v\,\big(\varphi_{r}'(v)^{2}+\tau\varphi_{r}(v)\big)\Big)\;.
\end{equation}
For current fluctuations, the upper limit $\nu_{r}$ of the integral in the action verifies $\varphi_{r}(\nu_{r})\propto\lambda$, and $\varphi$ can thus be interpreted as a field conjugate to the current. The integral is regularized at $-\infty$.

When $\tau\to\infty$, only the stationary eigenstate $r=0$ contributes to (\ref{functional integral}). When $\tau\to0$, the number of eigenstates $r$ contributing diverges, which makes it difficult to evaluate numerically our expressions (\ref{sigma average step})-(\ref{xi GF step}) for very small $\tau$. With the values of $\tau$ accessible numerically, we are however able to confirm that the various observables connect as expected to the results for TASEP on $\mathbb{Z}$. A proof is missing at the moment. The Fredholm determinant (\ref{Fredholm}), which is rather similar to known expressions for $F_{0}$, $F_{1}$, $F_{2}$, seems a good starting point.

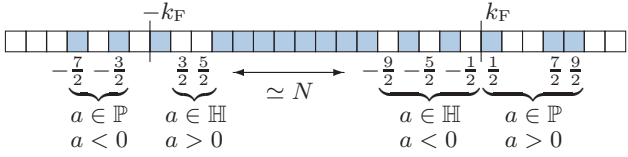
\begin{figure}
  \begin{center}
    \begin{tabular}{c}
      \setlength{\unitlength}{0.55mm}
      \begin{picture}(150,5)
        \put(50,0){\color[rgb]{0.7,0.8,0.9}\polygon*(0,0)(40,0)(40,5)(0,5)}
        \put(15,0){\color[rgb]{0.7,0.8,0.9}\polygon*(0,0)(5,0)(5,5)(0,5)}
        \put(25,0){\color[rgb]{0.7,0.8,0.9}\polygon*(0,0)(5,0)(5,5)(0,5)}
        \put(35,0){\color[rgb]{0.7,0.8,0.9}\polygon*(0,0)(5,0)(5,5)(0,5)}
        \put(95,0){\color[rgb]{0.7,0.8,0.9}\polygon*(0,0)(5,0)(5,5)(0,5)}
        \put(105,0){\color[rgb]{0.7,0.8,0.9}\polygon*(0,0)(5,0)(5,5)(0,5)}
        \put(115,0){\color[rgb]{0.7,0.8,0.9}\polygon*(0,0)(5,0)(5,5)(0,5)}
        \put(130,0){\color[rgb]{0.7,0.8,0.9}\polygon*(0,0)(10,0)(10,5)(0,5)}
        \multiput(0,0)(5,0){30}{\polygon(0,0)(5,0)(5,5)(0,5)}
        \put(33,8){$-k_{\text{F}}$}
        \put(116,8){$k_{\text{F}}$}
        \put(35,-2.5){\line(0,1){10}}
        \put(115,-2.5){\line(0,1){10}}
        \put(11,-6){$-\frac{7}{2}$}
        \put(21,-6){$-\frac{3}{2}$}
        \put(41,-6){$\frac{3}{2}$}
        \put(46,-6){$\frac{5}{2}$}
        \put(86,-6){$-\frac{9}{2}$}
        \put(96,-6){$-\frac{5}{2}$}
        \put(106,-6){$-\frac{1}{2}$}
        \put(116,-6){$\frac{1}{2}$}
        \put(131,-6){$\frac{7}{2}$}
        \put(136,-6){$\frac{9}{2}$}
        \put(15.5,-8){$\underbrace{\hspace{14\unitlength}}$}
        \put(40.5,-8){$\underbrace{\hspace{9\unitlength}}$}
        \put(90.5,-8){$\underbrace{\hspace{24\unitlength}}$}
        \put(115.5,-8){$\underbrace{\hspace{24\unitlength}}$}
        \put(15.7,-17){$a\in\mathbb{P}$}
        \put(15.7,-23){$a<0$}
        \put(38.7,-17){$a\in\mathbb{H}$}
        \put(38.7,-23){$a>0$}
        \put(95.3,-17){$a\in\mathbb{H}$}
        \put(95.3,-23){$a<0$}
        \put(120.7,-17){$a\in\mathbb{P}$}
        \put(120.7,-23){$a>0$}
        \put(55,-5){\vector(1,0){27}}
        \put(82,-5){\vector(-1,0){27}}
        \put(63,-11.5){$\simeq N$}
      \end{picture}\\\\
      \hspace*{0mm}
    \end{tabular}
  \end{center}
  \caption{Sets $\mathbb{P}$ and $\mathbb{H}$ for a choice of the $k_{j}$ (colored squares).}
  \label{Fig P H}
\end{figure}

\begin{figure}
  \begin{tabular}{ll}
    \begin{tabular}{l}\includegraphics[width=40mm]{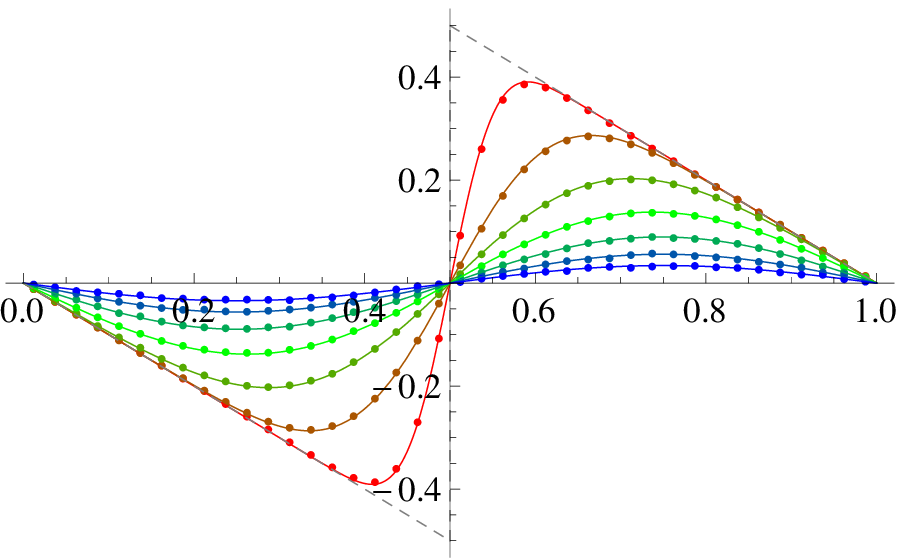}\end{tabular}
    \begin{picture}(0,0)
      \put(-8,11){(a)}
      \put(-41,11){$2\tau\langle\sigma(x,\tau)\rangle$}
      \put(-3,2){\scriptsize$x$}
    \end{picture}
    &
    \begin{tabular}{l}\includegraphics[width=40mm]{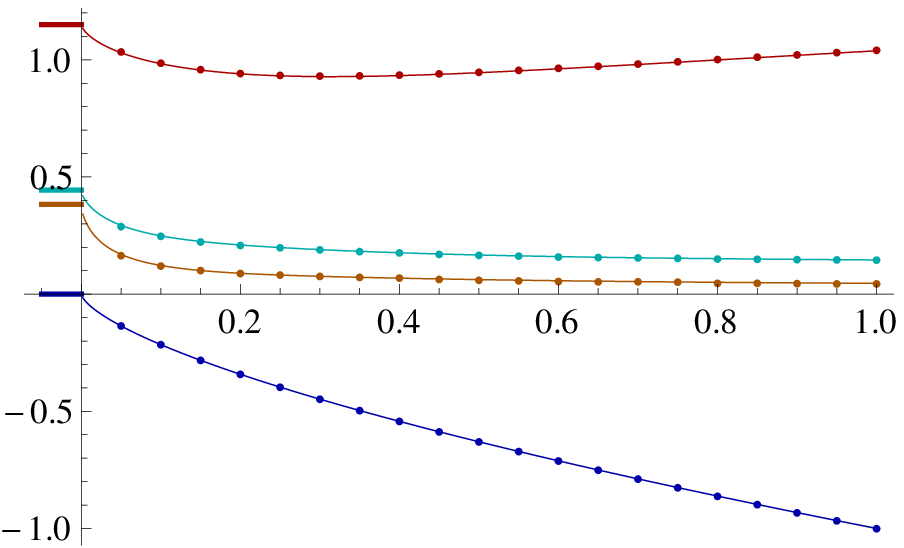}\end{tabular}
    \begin{picture}(0,0)
      \put(-7,6){(b)}
      \put(-25,5){$\langle\tilde{\xi}_{\tau}^{k}\rangle_{\text{c}}^{\text{stat}}$}
      \put(-2,1){\scriptsize$\tau$}
    \end{picture}
    \vspace{3mm}\\
    \begin{tabular}{l}\includegraphics[width=40mm]{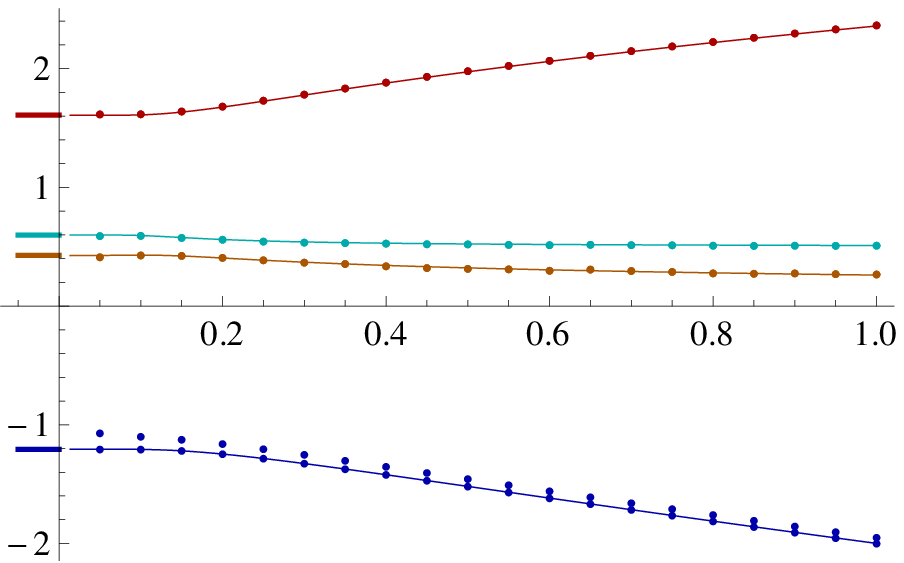}\end{tabular}
    \begin{picture}(0,0)
      \put(-38,11){(c)}
      \put(-15,6){$\langle\tilde{\xi}_{\tau}^{k}\rangle_{\text{c}}^{\text{flat}}$}
      \put(-2,0.7){\scriptsize$\tau$}
      \put(-36,-4.2){\tiny$\sim\!1\!/\!\sqrt{L}$}
      \put(-36.6,-6.5){\line(0,1){2}}
      \put(-37.1,-6.5){\line(1,0){1}}
      \put(-37.1,-4.5){\line(1,0){1}}
    \end{picture}
    &
    \begin{tabular}{l}\includegraphics[width=40mm]{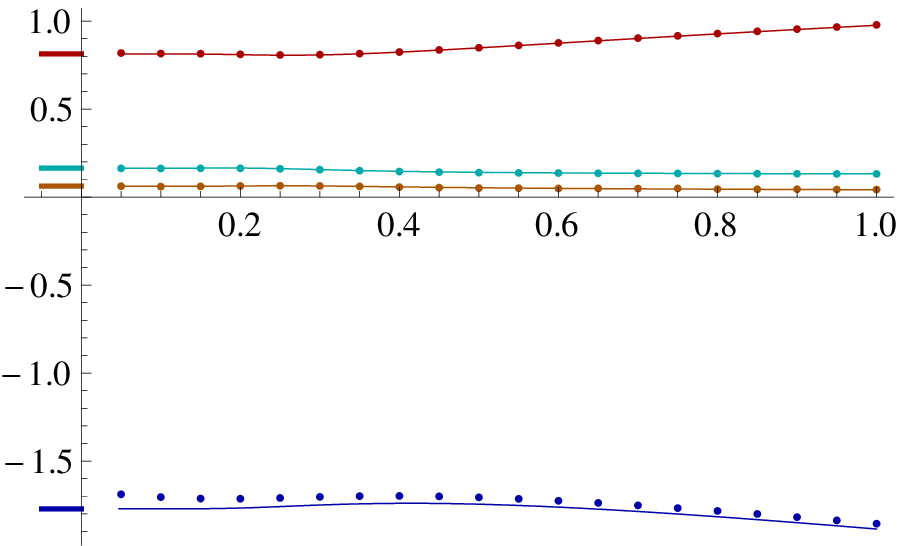}\end{tabular}
    \begin{picture}(0,0)
      \put(-7,-4){(d)}
      \put(-23,-3.5){$\langle\tilde{\xi}_{\tau}^{k}\rangle_{\text{c}}^{\text{step}}$}
      \put(-2,5.3){\scriptsize$\tau$}
      \put(-34.8,-7.4){\tiny$\sim\!1\!/\!\sqrt{L}$}
      \put(-35.6,-9.7){\line(0,1){2}}
      \put(-36.1,-9.7){\line(1,0){1}}
      \put(-36.1,-7.7){\line(1,0){1}}
    \end{picture}
  \end{tabular}
  \caption{(a) Average density profile for step initial condition at time $\tau=0.05$ (red curve, largest amplitude), $0.10$, \ldots, $0.35$ (blue curve, smallest amplitude). The solid lines correspond to a numerical evaluation of the exact formula (\ref{sigma average step}). The dots are the result of simulations. The dashed line is the ramp with a shock at $x=0.5$ obtained at small $\tau$ according to Burgers' equation. (b)-(d) Four first cumulants $\langle\tilde{\xi}_{\tau}^{k}\rangle_{\text{c}}$ of the rescaled current fluctuations for stationary (b), flat (c) and step (d) initial condition with $x=0$, from numerical evaluations of (\ref{xi GF stat})-(\ref{xi GF step}). From top to bottom are plotted the variance (red), third (cyan) and fourth (orange) cumulants and the mean value (blue). The marks on the left represent the short time limits $F_{0}$, $F_{1}$, $F_{2}$. The dots are the result of simulations. The two sequences for $\langle\tilde{\xi}_{\tau}\rangle$ in (c) correspond to $Q_{i}$ at $i$ either initially occupied (lower dots) or empty (upper dots).}
  \label{Fig aRho cQ}
\end{figure}

\begin{figure}
  \begin{tabular}{ll}
    \begin{tabular}{l}\includegraphics[width=40mm]{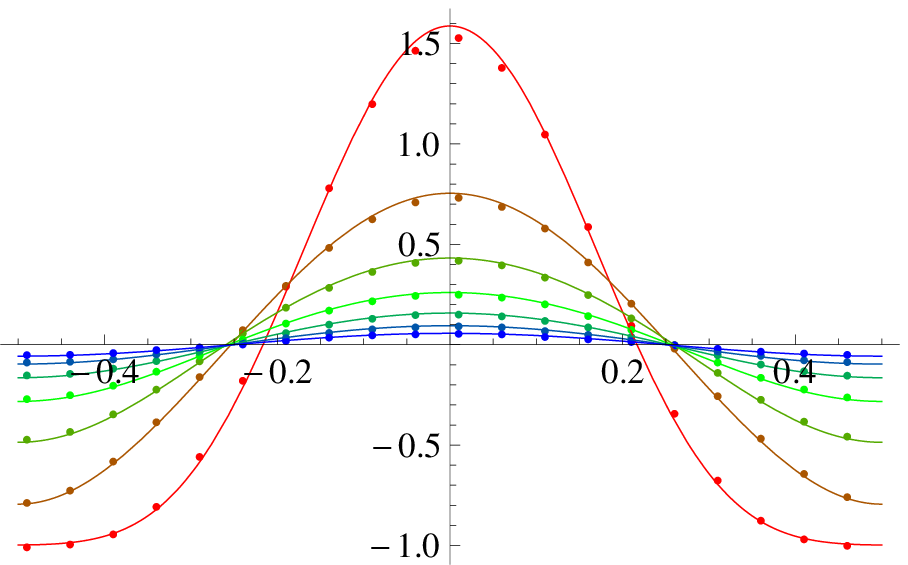}\end{tabular}
    \begin{picture}(0,0)
      \put(-7,11){(a)}
      \put(-40,11){$S(x,\tau)$}
      \put(-3.5,-0.5){\scriptsize$x$}
    \end{picture}
    &
    \begin{tabular}{l}\includegraphics[width=40mm]{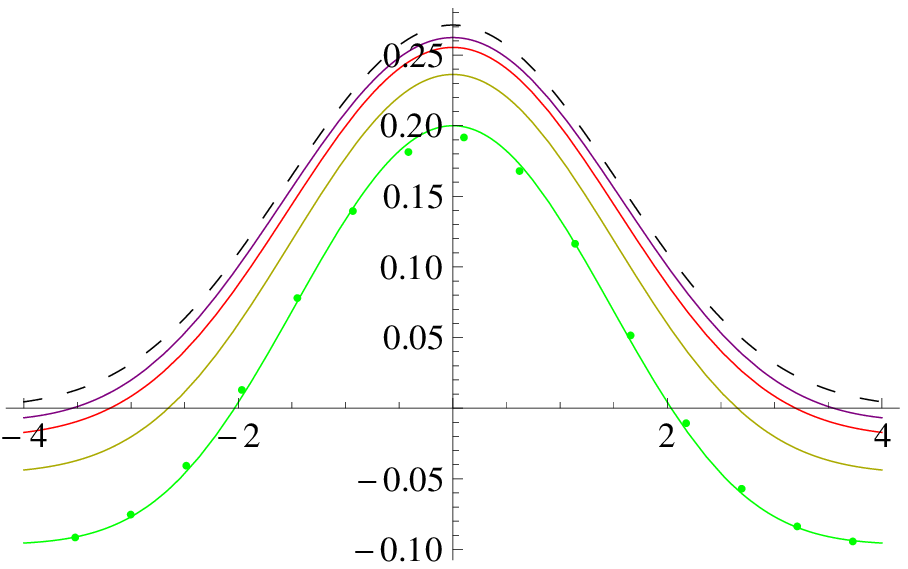}\end{tabular}
    \begin{picture}(0,0)
      \put(-7,11){(b)}
      \put(-42.5,11.5){\scriptsize$\tau^{2/3}\hspace{-1pt}S\hspace{-1pt}(\hspace{-1pt}\tau^{2/3}\hspace{-1pt}y,\hspace{-1pt}\tau\hspace{-1pt})$}
      \put(-3.5,-2.5){\scriptsize$y$}
    \end{picture}
  \end{tabular}
  \caption{(a) Stationary two-point function of the density $S(x,\tau)$ plotted as a function of $x$ for time $\tau=0.04$ (red curve, largest amplitude), $0.08$, \ldots, $0.28$ (blue curve, smallest amplitude). The dots are the result of simulations. (b) Plots of $\tau^{2/3}S(\tau^{2/3}y,\tau)$ as a function of $y$ for smaller times $\tau=0.001$ (upper, solid purple curve), $0.003$, $0.01$, $0.03$ (lower, solid green curve). The dots for $\tau=0.03$ are the result of simulations. The black, dashed curve is the Pr\"ahofer-Spohn scaling function $f_{\text{KPZ}}(y/2)/2$, plotted from \cite{PrahoferWeb}.}
  \label{Fig 2ptRho}
\end{figure}

\begin{figure}
  \begin{tabular}{ll}
    \begin{tabular}{l}\includegraphics[width=40mm]{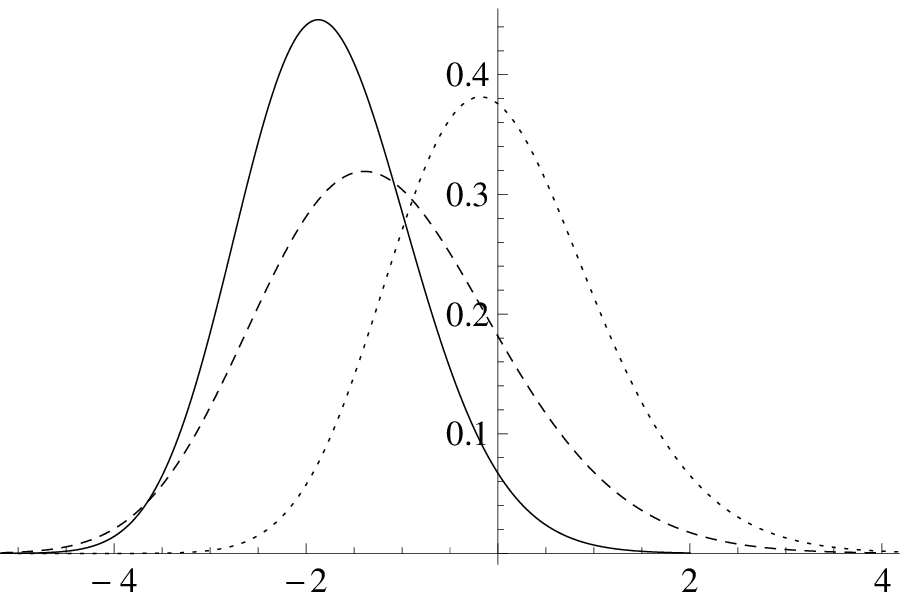}\end{tabular}
    \begin{picture}(0,0)
      \put(-39,11){(a)}
      \put(-34.7,7){$F_{2}'$}
      \put(-31,-2){$F_{1}'$}
      \put(-15,1){$F_{0}'$}
    \end{picture}
    &
    \begin{tabular}{l}\includegraphics[width=40mm]{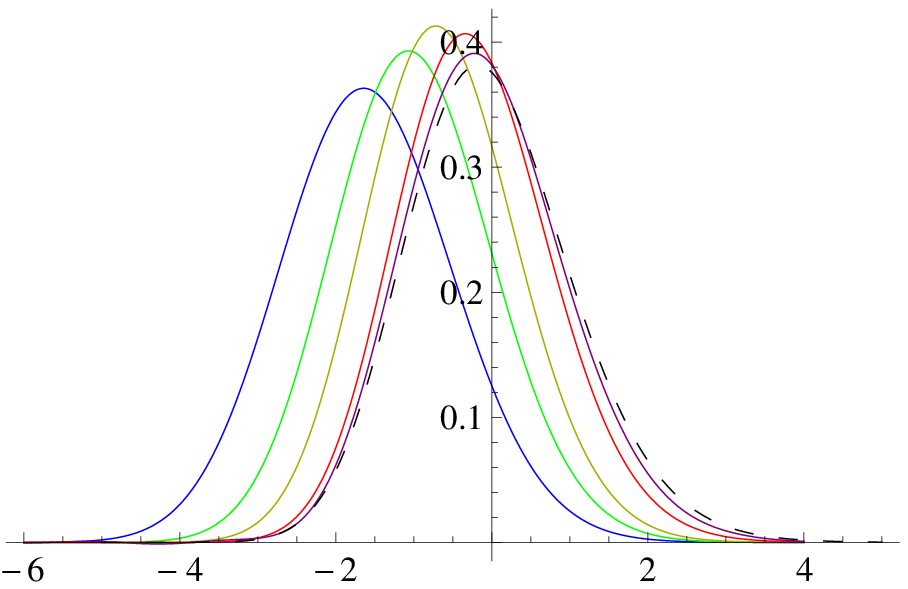}\end{tabular}
    \begin{picture}(0,0)
      \put(-39,11){(b)}
      \put(-15,11){$\tilde{P}_{\tau}^{\text{stat}}$}
    \end{picture}
    \vspace{3mm}\\
    \begin{tabular}{l}\includegraphics[width=40mm]{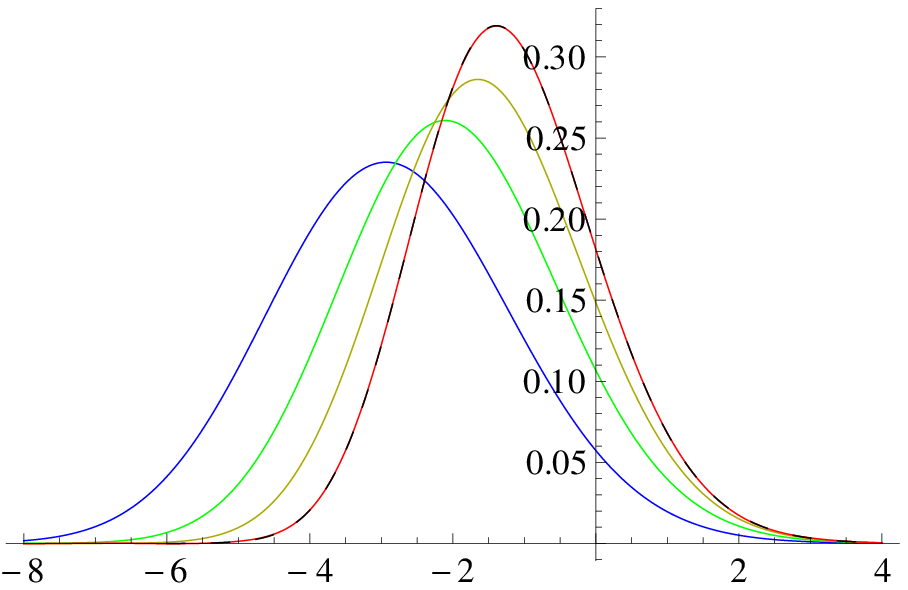}\end{tabular}
    \begin{picture}(0,0)
      \put(-39,11){(c)}
      \put(-12,11){$\tilde{P}_{\tau}^{\text{flat}}$}
    \end{picture}
    &
    \begin{tabular}{l}\includegraphics[width=40mm]{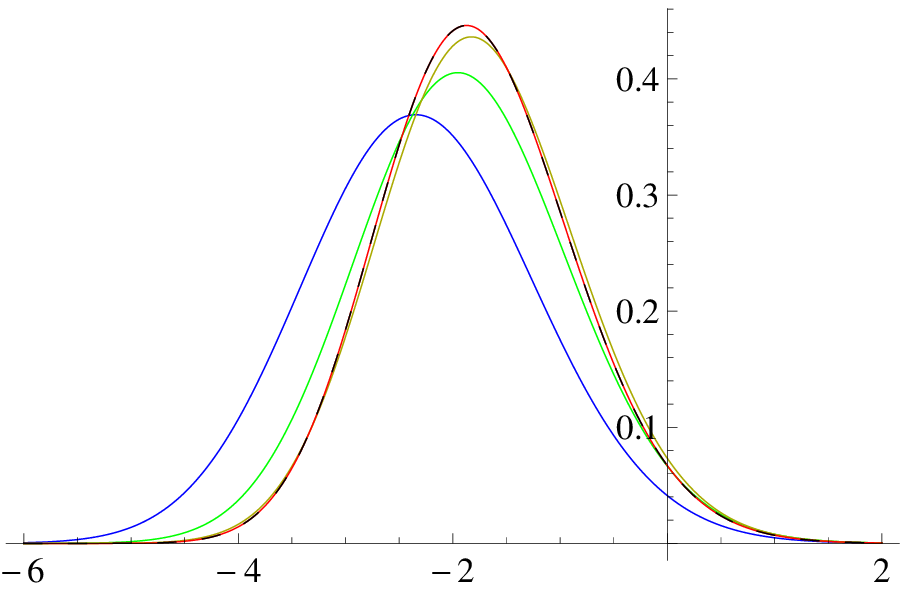}\end{tabular}
    \begin{picture}(0,0)
      \put(-39,11){(d)}
      \put(-10,11){$\tilde{P}_{\tau}^{\text{step}}$}
    \end{picture}
  \end{tabular}
  \caption{(a) Probability density of Baik-Rains distribution $F_{0}$ (dotted line), GOE and GUE Tracy-Widom distributions $F_{1}$ (dashed line) and $F_{2}$ (solid line), evaluated from \cite{PrahoferWeb}. (b)-(d) Probability density $\tilde{P}_{\tau}$ of rescaled current fluctuations $\tilde{\xi}_{\tau}$ for stationary (b), flat (c), step initial condition at $x=0$ (d). The black, dashed curves are the short time limits $F_{0}$, $F_{1}$, $F_{2}$. The solid curves are finite time distributions at $\tau=2$ (lowest, blue curves), $1$, $0.5$, $0.1$ (red), and $0.02$ (purple) for (b).}
  \label{Fig PDF Qloc stat flat step}
\end{figure}

From Bethe ansatz, each eigenstate of TASEP is fully characterized by $N$ momenta of fermionic quasi-particles, $k_{j}$, $j=1,\ldots,N$, integers or half-integers depending on the parity of $N$ and distinct modulo $L$. The stationary eigenstate corresponds to the Fermi sea $k_{j}=j-(N+1)/2$ with Fermi momentum $k_{\text{F}}=N/2$, and each excited state $r>0$ contributing to the time scale $t\sim L^{3/2}$ is specified by two finite sets of half-integers $\mathbb{P},\mathbb{H}\subset\mathbb{Z}+\frac{1}{2}$ describing momenta of particles and hole excitations, see figure \ref{Fig P H} and supplemental material. Excitations on both sides of the Fermi sea are independent, thus both $\mathbb{P}$ and $-\mathbb{H}$ have $m_{r}=m_{+}+m_{-}$ elements, with $m_{\pm}$ the number of elements with sign $\pm$. The stationary state corresponds to $\mathbb{P}=\mathbb{H}=\emptyset$. The total momentum of an eigenstate is $p_{r}=\sum_{a\in\mathbb{P}}a-\sum_{a\in\mathbb{H}}a$.

The field $\varphi_{r}$ is the derivative $\chi_{r}'$ of $\chi_{r}$, defined by $\chi_{r}(v)=\chi_{0}(v)+\sum_{a\in\mathbb{P}}\omega_{a}^{3}(v)/3+\sum_{a\in\mathbb{H}}\omega_{a}^{3}(v)/3$ with $\chi_{0}(v)=-(2\pi)^{-1/2}\Li_{5/2}(-\rme^{v})$ a polylogarithm and $\omega_{a}(v)=2(\sgn(a)\rmi\pi)^{1/2}(|a|+\sgn(a)\frac{\rmi v}{2\pi})^{1/2}$ the excitation with momentum $a$. The linear potential and kinetic part of the action come respectively from asymptotics of eigenvalues \cite{GS1992.1,P2014.1} and eigenvectors \cite{P2015.2} of $M(\lambda)$. They appear in (\ref{sigma average step})-(\ref{xi GF step}) as $\tau\chi_{r}(\nu_{r})$ and $D_{r}(\nu)^{2}$ with
\begin{align}
\label{D1}
D_{r}(\nu)=&\frac{(\frac{\rmi\pi}{2})^{m_{r}^{2}}}{(2\pi)^{m_{r}}}\Big(\prod_{\substack{a,b\in\mathbb{P}\\a>b}}(a-b)\Big)\Big(\prod_{\substack{a,b\in\mathbb{H}\\a>b}}(a-b)\Big)\\
&\times\exp\Big(\lim_{\Lambda\to\infty}-m_{r}^{2}\log\Lambda+\int_{-\Lambda}^{\nu}\rmd v\,\frac{\chi_{r}''(v)^{2}}{2}\Big)\;,\nonumber
\end{align}
where $\log\Lambda$ cancels the divergence of the integral at $-\infty$.

We now state our main results. The density fluctuations $\displaystyle\sigma(x,\tau)=2\sqrt{L}(\eta_{i}-\tfrac{1}{2})$ at site $i=(\kappa+x)L$, for step initial condition $\mathcal{S}_{\kappa}$ with sites $\kappa L-N+1,\ldots,\kappa L$ occupied initially, are equal on average to
\begin{equation}
\label{sigma average step}
\langle\sigma(x,\tau)\rangle_{\text{step}}=-2\rmi\pi\sum_{r>0}\rme^{2\rmi\pi p_{r}x}\,\frac{p_{r}D_{r}^{2}(\nu_{r})\,\rme^{\tau\chi_{r}(\nu_{r})}}{\chi_{r}''(\nu_{r})}\;,
\end{equation}
where $\nu_{r}$ is the solution of $\chi_{r}'(\nu_{r})=0$, and the summation is over all eigenstates except the stationary state. The function $\langle\sigma(x,\tau)\rangle_{\text{step}}$ is plotted in figure \ref{Fig aRho cQ}a along with results from simulations for a system of $L=1000$ sites averaged over $10^{7}$ independent realizations. The agreement is excellent: the $1/L$ finite size corrections can hardly be seen. At short time, the density converges to the ramp centered at $x=0$ predicted from Burgers' equation. For stationary and flat initial condition, translation invariance implies $\langle\sigma(x,\tau)\rangle_{\text{flat/stat}}=0$.

Similarly, the stationary two-point correlation function $S(x,\tau)=\langle\sigma(0,0)\sigma(x,\tau)\rangle_{\text{stat}}$ for the density fluctuations at site $0$, time $0$ and site $i=xL$, time $t$ is equal to
\begin{equation}
\label{sigma 2pt stat}
S(x,\tau)=-(2\pi)^{5/2}\sum_{r>0}\rme^{2\rmi\pi p_{r}x}\,\frac{p_{r}^{2}D_{r}^{2}(\nu_{r})\,\rme^{\tau\chi_{r}(\nu_{r})}}{\rme^{\nu_{r}}\chi_{r}''(\nu_{r})}\;,
\end{equation}
with $\nu_{r}$ again solution of $\chi_{r}'(\nu_{r})=0$. The two-point function is plotted in figure \ref{Fig 2ptRho}a, along with results of simulations for a system of $L=100$ sites averaged over $10^{9}$ realizations. The agreement is good, except at very short time when the scaling $t\sim L^{3/2}$ is not well verified for $L=100$. At short time $\tau$, our numerics seem to confirm the expected limit $\tau^{2/3}S(\tau^{2/3}y,\tau)\to f_{\text{KPZ}}(y/2)/2$, see figure \ref{Fig 2ptRho}b. In contrast with the infinite line case, where the two-point function converges to $0$ when the two points are far away, $S(x,\tau)\to-1$ when $x\neq0$ and $\tau\to0$ since the stationary measure of the periodic model is not exactly a product measure.

We define current fluctuations as $\xi_{\tau}=\frac{Q_{i}-t/4-\mathcal{R}L}{\sqrt{L}/2}$. The constant $\mathcal{R}$, which comes from the integration of Burgers' current between time $0$ and infinity \cite{P2015.3}, is equal to $0$ for stationary and flat initial condition, and the generating function of $\xi_{\tau}$ is given respectively by
\begin{equation}
\label{xi GF stat}
\langle\rme^{s\xi_{\tau}}\rangle_{\text{stat}}=\sqrt{2\pi}s^{2}\sum_{r}\frac{D_{r}^{2}(\nu_{r})\,\rme^{\tau\chi_{r}(\nu_{r})}}{\rme^{\nu_{r}}\chi_{r}''(\nu_{r})}
\end{equation}
and
\begin{equation}
\label{xi GF flat}
\langle\rme^{s\xi_{\tau}}\rangle_{\text{flat}}=s\sum_{r}\openone_{\{\mathbb{P}=\mathbb{H}\}}\frac{\rmi^{m_{r}}D_{r}(\nu_{r})\,\rme^{\tau\chi_{r}(\nu_{r})}}{\rme^{\nu_{r}/4}(1+\rme^{-\nu_{r}})^{1/4}\chi_{r}''(\nu_{r})}\;,
\end{equation}
with $\nu_{r}$ now solution of $\chi_{r}'(\nu_{r})=s$. For step initial condition $\mathcal{S}_{\kappa}$ with current counted at site $i=(\kappa+x)L$ modulo $L$ with $-1/2\leq x\leq1/2$, one has instead $\mathcal{R}=-|x|/2$ and
\begin{equation}
\label{xi GF step}
\langle\rme^{s\xi_{\tau}}\rangle_{\text{step}}=s\sum_{r}\rme^{2\rmi\pi p_{r}x}\,\frac{D_{r}^{2}(\nu_{r})\,\rme^{\tau\chi_{r}(\nu_{r})}}{\chi_{r}''(\nu_{r})}\;.
\end{equation}
Comparing (\ref{sigma average step}) and (\ref{xi GF step}) gives the conservation law $\langle\partial_{\tau}\sigma\rangle+\langle\partial_{x}j\rangle=0$ with instantaneous current $j=\partial_{\tau}\xi_{\tau}$.

The first cumulants of $\xi_{\tau}$, obtained from the generating functions (\ref{xi GF stat})-(\ref{xi GF step}), are plotted in figures \ref{Fig aRho cQ}b-\ref{Fig aRho cQ}d along with results of simulations for a system of $L=1000$ sites averaged over $10^{7}$ realizations. The agreement is very good. We checked from simulations of smaller systems that the discrepancies for the mean value with step and flat initial condition are due to finite size corrections of order $1/\sqrt{L}$ instead of $1/L$ in all the other cases.

The stationary large deviations are reached at long time from the contribution $\rme^{\tau\chi_{0}(\nu_{0})}$ of the stationary state. The results on the infinite line are recovered numerically at short time: defining $\tilde{\xi}_{\tau}=-\xi_{\tau}/\tau^{1/3}$ for stationary initial condition, $\tilde{\xi}_{\tau}=-2^{2/3}\xi_{\tau}/\tau^{1/3}$ for flat initial condition and $\tilde{\xi}_{\tau}=-(\xi_{\tau}-\frac{x^{2}}{4\tau})/\tau^{1/3}$ for step initial condition, the cumulants of $\tilde{\xi}_{\tau}$ have a finite limit at small $\tau$. They converge respectively to the cumulants of $F_{0}$, $F_{1}$, and $F_{2}$ (with $x\neq1/2$; numerics indicate a different distribution around the position of the shock $x=1/2$ in the step case).

The probability density $P_{\tau}$ of $\xi_{\tau}$, extracted from the generating function by Fourier transform $P_{\tau}(u)=\int_{-\infty}^{\infty}\frac{\rmd s}{2\pi}\,\rme^{-\rmi su}\langle\rme^{\rmi s\xi_{\tau}}\rangle$, is plotted under rescaling by $t^{1/3}$ in figures \ref{Fig PDF Qloc stat flat step}b-\ref{Fig PDF Qloc stat flat step}d.
After some manipulations described in the supplemental material, the cumulative distribution function $F_{\tau}(u)=P(\xi_{\tau}>u)$ reduces in all cases to integrals of a Fredholm determinant,
\begin{equation}
\label{Fredholm}
F_{\tau}(u)=\int\frac{\rmd\nu}{2\rmi\pi}\,\alpha(\nu)\,\rme^{\tau\chi_{0}(\nu)-u\chi_{0}'(\nu)}\oint\!\frac{\rmd z}{2\rmi\pi z}\,\det(\openone-J)\;.
\end{equation}
The integral over $\nu$ goes from $\rme^{-\rmi\theta}\infty$ to $\rme^{\rmi\theta}\infty$ with $\pi/5<\theta<\pi/2$. The function $\alpha$ is respectively equal for stationary, flat and step initial condition to $\alpha_{\text{stat}}(\nu)=-\sqrt{2\pi}\rme^{-\nu}D_{0}^{2}(\nu)\partial_{u}$, $\alpha_{\text{flat}}(\nu)=D_{0}(\nu)\,\rme^{-\nu/4}/(1+\rme^{-\nu})^{1/4}$, and $\alpha_{\text{step}}(\nu)=D_{0}(\nu)^{2}$. The operator $J$ is defined from $K$, $L$ and $X$ acting on $\mathbb{Z}+1/2$, with kernels $K_{a,b}=\delta_{a,b}\,\exp\big(\frac{\tau\,\omega_{a}^{3}(\nu)}{3}+u\,\omega_{a}(\nu)\big)$, $L_{a,b}=z^{\sgn a}f_{a}(\nu)f_{b}(\nu)/(\omega_{a}(\nu)+\omega_{b}(\nu))$ with $f_{a}(\nu)=\omega_{a}(\nu)^{-1/2}\exp\big(\int_{-\infty}^{\nu}\rmd v\,\frac{\chi_{0}''(v)}{\omega_{a}(v)}\big)$, and $X_{a,b}=\delta_{a,b}\,\rme^{2\rmi\pi a x}$. One has $J_{\text{stat}}=(KL)^{2}$, $J_{\text{flat}}=K^{2}L$ and $J_{\text{step}}=XKLX^{-1}KL$. For an evolution conditioned on flat initial and final states, the Fredholm determinant is replaced by the infinite product $\prod_{a\in\mathbb{Z}+1/2}(1+z^{\sgn a}K_{a,a}^{2})$ \cite{P2015.3}, and leads to large deviations at short time similar to the ones observed numerically in \cite{LK2006.1} for the total current.

\textit{Conclusions}. The exact formulas derived in this letter for TASEP using the special structure of its Bethe ansatz generalize several known results for the infinite system and the non-equilibrium steady state of the periodic model. They provide the full crossover on the relaxation time scale $t\sim L^{3/2}$. Our main results (\ref{sigma average step})-(\ref{xi GF step}) are expressed in a unified way and point to a field theoretic description of KPZ fluctuations in terms of a scalar field in a linear potential, $\varphi$, conjugate to the current.

Our results should hold for any model of one-dimensional KPZ universality. It might be possible to recover them in other exactly solvable models, in particular stochastic Burgers' equation, using the replica method with asymptotics of eigenstates for the attractive $\delta$-Bose gas in finite volume, and ASEP, a generalization of TASEP where particles also hop backward with rate $q<1$, related to a twisted XXZ spin chain with anisotropy $\Delta=(q^{1/2}+q^{-1/2})/2>1$.

It would be interesting to generalize our results to the full statistics of time-dependent density and current profiles, and understand how the field $\varphi$ couples to an arbitrary initial condition $\sigma(x,0)$ beyond the three specific initial states studied here. Another interesting but difficult question is the extension to the crossover between the far from equilibrium KPZ fluctuations and the Edwards-Wilkinson fluctuations of an interface at equilibrium, corresponding for ASEP to the weakly asymmetric regime $q\to1$ with $1-q\sim1/\sqrt{L}$.

\pagebreak
\begin{center}
\textbf{\large Supplemental material}
\end{center}
\setcounter{equation}{0}
\setcounter{figure}{0}
\setcounter{table}{0}
\setcounter{page}{1}
\makeatletter
\renewcommand{\theequation}{S\arabic{equation}}
\renewcommand{\thefigure}{S\arabic{figure}}
\renewcommand{\bibnumfmt}[1]{[S#1]}
\makeatother
\renewcommand{\cite}[1]{[S\citenum{#1}]}
\newcommand{\citeTwo}[2]{[S\citenum{#1},\,S\citenum{#2}]}

In this supplemental material, we briefly describe in the first section some aspects of the Bethe ansatz for TASEP, in particular asymptotics of eigenstates, some of them being new. In the second section, we derive the Fredholm determinant expression (\ref{Fredholm}) for the cumulative distribution function of current fluctuations.

\begin{section}{Large \texorpdfstring{$L$}{L} asymptotics of TASEP eigenstates from Bethe ansatz}
The Bethe ansatz gives each eigenvector of the deformed Markov matrix $M(\lambda)$ of TASEP with $N$ particles on a periodic lattice of $L$ sites as a linear combination of plane waves with $N$ complex momenta $q_{j}$. It reduces for TASEP to the determinant
\begin{equation}
\langle\mathcal{C}|\phi\rangle\propto\det\big(\rme^{\rmi q_{j}x_{\ell}}/(1-\rme^{\rmi q_{j}-\lambda/L})^{\ell}\big)_{j,\ell}\;,
\end{equation}
with particles at positions $1\leq x_{1}<\ldots<x_{N}\leq L$. Since particles are confined inside a box, the momenta take discrete values, solution of the Bethe equations
\begin{equation}
\label{Bethe equations}
\rme^{\rmi q_{j}}\Big(\frac{[1-\rme^{\rmi q-\lambda/L}]}{1-\rme^{\rmi q_{j}-\lambda/L}}\Big)^{N/L}=\rme^{2\rmi\pi k_{j}/L}\;.
\end{equation}
The set of numbers $k_{j}$, integers or half-integers depending on the parity of $N$ and distinct modulo $L$, fully characterizes the eigenstate. The geometric mean $[1-\rme^{\rmi q-\lambda/L}]=\prod_{j=1}^{N}(1-\rme^{\rmi q_{j}-\lambda/L})^{1/N}$ induces correlations between the momenta. The corresponding eigenvalue of $M(\lambda)$ is $E=\sum_{j=1}^{N}(\rme^{-\rmi q_{j}+\lambda/L}-1)$, and the one of the translation operator is $\rme^{2\rmi\pi p/L}$ with total momentum $p=\sum_{j=1}^{N}k_{j}$.

Bethe ansatz naturally gives finite size expressions for the eigenstates. The main technical difficulty is usually to take the thermodynamic limit $L,N\to\infty$, which is needed in order to derive the main results (\ref{sigma average step})-(\ref{xi GF step}). A first step was the exact calculation \cite{S_GS1992.1} of the gap of the Markov matrix $M$, shown to scale as $L^{-3/2}$. A key observation is that lower eigenstates can be understood by a particle-hole picture, where the $k_{j}$'s are interpreted as momenta of fermionic quasiparticles without spin. The stationary eigenstate corresponds to the Fermi sea $\{k_{j}^{0}=j-(N+1)/2$, $j=1,\ldots,N\}$. Eigenstates close to it are obtained by exciting quasiparticles of momenta $k^{0}$, $|k^{0}|<k_{\text{F}}$ to $k$, $|k|>k_{\text{F}}$ with Fermi momentum $k_{\text{F}}=N/2$. This creates vacancies in the Fermi sea, viewed as holes of momenta $-k^{0}$. It is conjectured \cite{S_P2014.1} based on numerics for small systems that all the eigenstates of $M(\lambda)-\lambda/4\,\openone$ with an eigenvalue of order $L^{-3/2}$ are generated by particle-hole excitations at a finite distance of the Fermi surface $\pm k_{\text{F}}$, characterized by the two finite sets of half-integers $\mathbb{P},\mathbb{H}\subset\mathbb{Z}+1/2$ of figure \ref{Fig P H}.

The various known asymptotics of eigenvalues \cite{S_P2014.1} and eigenvectors \citeTwo{S_P2015.2}{S_P2015.3} of $M(\lambda)$ with finite rescaled fugacity
\begin{equation}
s=\lambda\sqrt{L}/2\;
\end{equation}
involve the elementary excitation $\omega_{a}(v)$ with momentum $a$, defined above (\ref{D1}), which verifies $\omega_{a}'(v)=-1/\omega_{a}(v)$ and is analytic for $v$ in the domain $\mathbb{D}=\mathbb{C}\backslash\big(\rmi[\pi,\infty)\cup-\rmi[\pi,\infty)\big)$ when $a\in\mathbb{Z}+1/2$. From these elementary excitations, one introduces above (\ref{D1}) the function $\chi_{r}$ with $\chi_{0}(v)$ given in terms of Hurwitz $\zeta$ function by $\chi_{0}(v)=\frac{8\pi^{3/2}}{3}\big(\sqrt{-\rmi}\,\zeta\big(-\frac{3}{2},\frac{1}{2}+\frac{\rmi v}{2\pi}\big)+\sqrt{\rmi}\,\zeta\big(-\frac{3}{2},\frac{1}{2}-\frac{\rmi v}{2\pi}\big)\big)$, analytic in $\mathbb{D}$. For $\Re v<0$ and for $\Re v>0$, $-\pi<\Im v<\pi$, $\chi_{0}(v)$ reduces to the polylogarithm given above (\ref{D1}). One also defines the solution $\nu_{r}$ of $\chi_{r}'(\nu_{r})=s$ for $\Re s\geq0$, which induces a coupling between the quasiparticles. The quantity $\nu_{r}$ is singular for the stationary state when $s\to0$, with $\nu_{0}\to-\infty$ and $\chi_{0}''(\nu_{0})\simeq\rme^{\nu_{0}}/\sqrt{2\pi}$.

The Euler-Maclaurin formula allows to derive \cite{S_P2014.1} the large $L$ asymptotics $[1-\rme^{\rmi q-\lambda/L}]\simeq\frac{1}{4}+\frac{\nu_{r}}{2L}$ for the geometric mean of the Bethe roots, and the asymptotics
\begin{equation}
E_{r}-\frac{\lambda}{4}\simeq\frac{\chi_{r}(\nu_{r})}{2L^{3/2}}\;
\end{equation}
for the eigenvalue. A key point for the derivation is that the Bethe roots $\rme^{\rmi q_{j}}$ accumulate on a closed contour in $\mathbb{C}$, and Euler-Maclaurin's leading integral can be computed explicitly using residues.

The norm of the Bethe eigenstates of $M(\lambda)$, given in general by the Gaudin determinant, simplifies for TASEP \cite{S_MSS2012.2}. The components of the eigenvectors for flat and step configuration reduce to Vandermonde determinants, and have been computed in the thermodynamic limit \citeTwo{S_P2015.2}{S_P2015.3} using the Euler-Maclaurin formula. We use the same normalization of the eigenvectors $\phi_{r}$ as in \cite{S_P2015.3}. It verifies $\langle\mathcal{C}|\phi_{r}\rangle=\langle\phi_{r}|\tilde{\mathcal{C}}\rangle$, where the configurations $\mathcal{C}$ and $\tilde{\mathcal{C}}$, corresponding respectively to particles at positions $1\leq x_{1}<\ldots<x_{N}\leq L$ and $1\leq\tilde{x}_{1}<\ldots<\tilde{x}_{N}\leq L$, are related by space reversal $\tilde{x}_{j}=L+1-x_{N+1-j}$. The scalar product is then equal for large $L$ to
\begin{equation}
\frac{\Omega}{\langle\phi_{r}|\phi_{r}\rangle}\simeq\frac{\rme^{\nu_{r}}}{\sqrt{2\pi}\chi''(\nu_{r})}\;,
\end{equation}
where $\Omega=\C{L}{N}$ is the total number of configurations. The component of the left eigenvectors for a flat configuration $\mathcal{F}$ is
\begin{equation}
\langle\phi_{r}|\mathcal{F}\rangle\simeq\openone_{\{\mathbb{P}=\mathbb{H}\}}\frac{\rmi^{m_{r}}\,\rme^{-\nu_{r}/4}}{(1+\rme^{-\nu_{r}})^{1/4}}\;.
\end{equation}
For a step configuration $\mathcal{S}_{\kappa}$ with sites $\kappa L-N+1,\ldots,\kappa L$ occupied initially, one has
\begin{equation}
\langle\phi_{r}|\mathcal{S}_{\kappa}\rangle\simeq\rme^{-\lambda L/8}\,\rme^{-2\rmi\pi p_{r}\kappa}D_{r}(\nu_{r})\;,
\end{equation}
with $D_{r}$ defined in (\ref{D1}). All the expressions are understood as analytic in $\nu_{r}\in\mathbb{D}$. Note that there is a misprint in equation (35) of the published version of \cite{S_P2015.3}, with a global factor $\rmi^{m_{r}^{2}}$ missing.

The previous asymptotics are sufficient for studying an evolution conditioned on both initial and final configurations being flat or step \cite{S_P2015.3}. Relaxing the conditioning on the final state for the fluctuations of $Q_{i}$ requires the asymptotics of $\sum_{\mathcal{C}}\langle\mathcal{C}|\rme^{-\frac{\lambda}{L}\,S_{i}}|\phi_{r}\rangle$. It can be obtained from the finite size Bethe ansatz formula (valid without assuming that the $q_{j}$'s are solution of Bethe equations) $\sum_{\mathcal{C}}\langle\mathcal{C}|\rme^{-\frac{\lambda}{L}\,S_{0}}|\phi_{r}\rangle\propto\det V$ \cite{S_B2009.1} with $V_{j,k}=\sum_{\ell=0}^{j-1}(-1)^{\ell}\C{L}{\ell}y_{k}^{\ell+1-j}$ for $1\leq j<N$, $V_{N,k}=-\sum_{\ell=N-1}^{L}(-1)^{\ell}\C{L}{\ell}y_{k}^{\ell+1-N}$ and $y_{k}=1-\rme^{\rmi q_{k}-\lambda/L}$. The operator $S_{0}$ precisely cancels a factor from the eigenvector expressed in terms of the $y_{k}$'s, and makes the calculation of fluctuations possible for $Q_{i}$ but not for $Q$. The term $\ell=N-1$ can be removed from the sum for $V_{N,k}$ by linear combination with $V_{1,k}$. Rewriting $V_{N,k}=-y_{k}^{1-j}(1-y_{k})^{L}+\sum_{\ell=0}^{N-1}(-1)^{\ell}\C{L}{\ell}y_{k}^{\ell+1-N}$, expanding the row $j=N$ of the determinant, and doing further linear combinations to keep only the largest value of $\ell$ in each sum, $\det V$ reduces to a $N\times N$ Vandermonde determinant plus $N$ Vandermonde determinants of size $N-1$. Factorising everything leads to $\det V\propto\big(\prod_{j<k}(y_{j}-y_{k})\big)\Big(1-\sum_{j}(1-y_{j})^{L}\prod_{k\neq j}\frac{y_{k}}{y_{k}-y_{j}}\Big)$. Assuming now that the $y_{j}$'s are solution of the Bethe equations, the second factor reduces to $1-\rme^{-\lambda}$ after using the identity $\sum_{j=1}^{N}\prod_{k\neq j}\frac{y_{j}}{y_{j}-y_{k}}=1$. The asymptotics \cite{S_P2015.2} of the Vandermonde determinant in front leads to
\begin{equation}
\frac{1}{\Omega}\sum_{\mathcal{C}}\langle\mathcal{C}|\rme^{-\frac{\lambda}{L}\,S_{i}}|\phi_{r}\rangle\simeq\rme^{2\rmi\pi p_{r}i/L}\sqrt{2\pi}\,s\,\rme^{-\nu_{r}}D_{r}(\nu_{r})\;.
\end{equation}
Exactly the same asymptotics is found for $\langle\phi_{r}|\rme^{\frac{\lambda}{L}\,S_{-i}}|P_{\text{st}}\rangle$ with $|P_{\text{st}}\rangle=\frac{1}{\Omega}\sum_{\mathcal{C}}|\mathcal{C}\rangle$ the stationary state.

Inserting the operator $\openone_{\{\eta_{i}(\mathcal{C})=0\}}$ replaces $\det V$ by the determinant of another matrix $\tilde{V}$ \cite{S_B2009.1}, defined by replacing $L$ by $L-1$ everywhere in the expression of $V$. One finds then $\det\tilde{V}\propto\big(\prod_{j<k}(y_{j}-y_{k})\big)\Big(1-\sum_{j}(1-y_{j})^{L-1}\prod_{k\neq j}\frac{y_{k}}{y_{k}-y_{j}}\Big)$. Using the identity $\sum_{j=1}^{N}\frac{1}{1-y_{j}}\prod_{k\neq j}\frac{y_{j}}{y_{j}-y_{k}}=\prod_{j=1}^{N}\frac{1}{1-y_{j}}$, the Bethe equations imply that the second factor is equal to $1-\rme^{-L\gamma}\prod_{j=1}^{N}\frac{1}{1-y_{j}}$, and one has
\begin{equation}
\frac{1}{\Omega}\sum_{\mathcal{C}}\openone_{\{\eta_{i}(\mathcal{C})=0\}}\langle\mathcal{C}|\phi_{r}\rangle\simeq\frac{\sqrt{2\pi}}{2\sqrt{L}}\,\rme^{2\rmi\pi p_{r}i/L}2\rmi\pi p_{r}\,\rme^{-\nu_{r}}D_{r}(\nu_{r})
\end{equation}
at $\lambda=0$ for $r\neq0$. In the stationary state, the right hand side is replaced by $1/2$. The asymptotics of $\frac{1}{\Omega}\sum_{\mathcal{C}}\openone_{\{\eta_{-i}(\mathcal{C})=0\}}\langle\phi_{r}|\mathcal{C}\rangle$ is exactly the same.

Gathering all the large $L$ asymptotics above, one recovers the main results (\ref{sigma average step})-(\ref{xi GF step}) of the letter.
\end{section}

\begin{section}{Fredholm determinant for the current fluctuations}
The Fredholm determinant (\ref{Fredholm}) for the cumulative distribution function of current fluctuations is a consequence of an alternative expression for $D_{r}(\nu)$, obtained from (\ref{D1}) by computing explicitly the terms of the integral independent of $\chi_{0}$ using the identity $\partial_{v}\log(\omega_{a}(v)+\omega_{b}(v))=-\omega_{a}^{-1}(v)\omega_{b}^{-1}(v)$. It leads to the Cauchy determinant
\begin{equation}
\label{D2}
D_{r}(\nu)=\exp\Big(\int_{-\infty}^{\nu}\!\!\rmd v\,\frac{\chi_{0}''(v)^{2}}{2}\Big)\,\det\Big(\frac{\rmi\,f_{a}(\nu)f_{b}(\nu)}{\omega_{a}(\nu)+\omega_{b}(\nu)}\Big)_{\substack{a\in\mathbb{P}\\b\in\mathbb{H}}}\;
\end{equation}
with $f_{a}(\nu)=\omega_{a}(\nu)^{-1/2}\exp(\int_{-\infty}^{\nu}\rmd v\,\frac{\chi_{0}''(v)}{\omega_{a}(v)})$.

We observe that the Jacobian of the change of variables $s\to\nu=\nu_{r}$ in the integral for the probability density $P_{\tau}(u)=\int_{-\infty}^{\infty}\frac{\rmd s}{2\pi}\,\rme^{-\rmi su}\langle\rme^{\rmi s\xi_{\tau}}\rangle$ cancels the factor $\chi_{r}''(\nu_{r})$ in the denominator of the expressions (\ref{xi GF stat})-(\ref{xi GF step}) for $\langle\rme^{\rmi s\xi_{\tau}}\rangle$.

This allows to perform explicitly the summation over the eigenstates $r=(\mathbb{P},\mathbb{H})$ after introducing a contour integral to enforce the constraint that the number of positive elements of $\mathbb{P}$ minus the number of negative elements of $\mathbb{P}$ is equal to the number of negative elements of $\mathbb{H}$ minus the number of positive elements of $\mathbb{H}$. In the stationary and the step case, an extra step is needed to perform the summation over $\mathbb{H}$ and reduce the determinant squared to a single determinant, using a version of the Cauchy-Binet formula,
\begin{align}
\sum_{\mathbb{H}\subset\Gamma}\openone_{\{|\mathbb{H}|=|\mathbb{P}|\}}&\det(L_{a,b})_{\substack{a\in\mathbb{P}\\b\in\mathbb{H}}}\det(M_{b,c})_{\substack{b\in\mathbb{H}\\c\in\mathbb{P}}}\nonumber\\
&=\det\Big(\sum_{b\in\Gamma}L_{a,b}M_{b,c}\Big)_{\substack{a\in\mathbb{P}\\c\in\mathbb{P}}}\;.
\end{align}
The arbitrary set $\Gamma$ is equal to $\mathbb{Z}+1/2$ here. A Fredholm determinant is obtained in the end using the identity
\begin{equation}
\sum_{\mathbb{P}\subset\Gamma}\det(J_{a,b})_{\substack{a\in\mathbb{P}\\b\in\mathbb{P}}}=\det(\openone+J)\;,
\end{equation}
where $J$ is an operator acting on $\Gamma$, with discrete kernel $J_{a,b}$. One recovers (\ref{Fredholm}).

The expansion of the Fredholm determinant of (\ref{Fredholm})
\begin{equation}
\det(\openone-J)=\sum_{n=0}^{\infty}\frac{(-1)^{n}}{n!}\!\!\!\sum_{a_{1},\ldots,a_{n}\in\mathbb{Z}+\frac{1}{2}}\!\!\!\det(J_{a_{j},a_{k}})_{j,k}\;
\end{equation}
leads to a corresponding expansion of the cumulative distribution function of the current, $F_{\tau}(u)=\sum_{n=0}^{\infty}\mu_{n}(\tau,u)$. Numerics in the flat case indicate that the coefficients $\mu_{n}(\tau,\tau^{1/3}u)$ do not converge individually when $\tau\to0$: only the full sum $F_{\tau}(\tau^{1/3}u)$ has a finite limit, the GOE Tracy-Widom distribution $F_{1}$. This is in contrast with a common situation for long time limits of finite time Fredholm determinant expressions for TASEP on $\mathbb{Z}$, where all the terms of the expansion converge individually and it is sufficient to take the limit of the kernel.
\end{section}


\begin{thebibliography}{10}

\bibitem{KPZ1986.1}
M.~Kardar, G.~Parisi, and Y.-C. Zhang.
\newblock Dynamic scaling of growing interfaces.
\newblock {\em Phys. Rev. Lett.}, 56:889--892, 1986.

\bibitem{KK2010.1}
T.~Kriecherbauer and J.~Krug.
\newblock A pedestrian's view on interacting particle systems, {KPZ} universality and random matrices.
\newblock {\em J. Phys. A: Math. Theor.}, 43:403001, 2010.

\bibitem{QS2015.1}
J.~Quastel and H.~Spohn.
\newblock The one-dimensional {KPZ} equation and its universality class.
\newblock {\em J. Stat. Phys.}, 160:965--984, 2015.

\bibitem{HHT2015.1}
T.~Halpin-Healy and K.A. Takeuchi.
\newblock A {KPZ} cocktail-shaken, not stirred...
\newblock {\em J. Stat. Phys.}, 160:794--814, 2015.

\bibitem{MWIRMSM1998.1}
M.~Matsushita, J.~Wakita, H.~Itoh, I.~R\`afols, T.~Matsuyama, H.~Sakaguchi, and M.~Mimura.
\newblock Interface growth and pattern formation in bacterial colonies.
\newblock {\em Physica A}, 249:517--524, 1998.

\bibitem{MMAAMT2001.1}
M.~Myllys, J.~Maunuksela, M.~Alava, T.~Ala-Nissila, J.~Merikoski, and J.~Timonen.
\newblock Kinetic roughening in slow combustion of paper.
\newblock {\em Phys. Rev. E}, 64:036101, 2001.

\bibitem{TS2010.1}
K.A. Takeuchi and M.~Sano.
\newblock Universal fluctuations of growing interfaces: Evidence in turbulent liquid crystals.
\newblock {\em Phys. Rev. Lett.}, 104:230601, 2010.

\bibitem{TSSS2011.1}
K.A. Takeuchi, M.~Sano, T.~Sasamoto, and H.~Spohn.
\newblock Growing interfaces uncover universal fluctuations behind scale invariance.
\newblock {\em Sci. Rep.}, 1:34, 2011.

\bibitem{LK2006.1}
D.S. Lee and D.~Kim.
\newblock Universal fluctuation of the average height in the early-time regime of one-dimensional {K}ardar-{P}arisi-{Z}hang-type growth.
\newblock {\em J. Stat. Mech.}, 2006:P08014, 2006.

\bibitem{T2012.1}
K.A. Takeuchi.
\newblock Statistics of circular interface fluctuations in an off-lattice {E}den model.
\newblock {\em J. Stat. Mech.}, 2012:P05007, 2012.

\bibitem{HHL2014.1}
T.~Halpin-Healy and Y.~Lin.
\newblock Universal aspects of curved, flat, and stationary-state {K}ardar-{P}arisi-{Z}hang statistics.
\newblock {\em Phys. Rev. E}, 89:010103(R), 2014.

\bibitem{D1998.1}
B.~Derrida.
\newblock An exactly soluble non-equilibrium system: the asymmetric simple exclusion process.
\newblock {\em Phys. Rep.}, 301:65--83, 1998.

\bibitem{GM2006.1}
O.~Golinelli and K.~Mallick.
\newblock The asymmetric simple exclusion process: an integrable model for non-equilibrium statistical mechanics.
\newblock {\em J. Phys. A: Math. Gen.}, 39:12679--12705, 2006.

\bibitem{CMZ2011.1}
T.~Chou, K.~Mallick, and R.K.P. Zia.
\newblock Non-equilibrium statistical mechanics: from a paradigmatic model to biological transport.
\newblock {\em Rep. Prog. Phys.}, 74:116601, 2011.

\bibitem{CSS2000.1}
D.~Chowdhury, L.~Santen, and A.~Schadschneider.
\newblock Statistical physics of vehicular traffic and some related systems.
\newblock {\em Phys. Rep.}, 329:199--329, 2000.

\bibitem{KvO2010.1}
T~Karzig and F.~von Oppen.
\newblock Signatures of critical full counting statistics in a quantum-dot chain.
\newblock {\em Phys. Rev. B}, 81:045317, 2010.

\bibitem{S2012.1}
H.~Spohn.
\newblock Stochastic integrability and the {KPZ} equation.
\newblock {\em IAMP news bulletin}, pages 5--9, April 2012.

\bibitem{C2011.1}
I.~Corwin.
\newblock The {K}ardar-{P}arisi-{Z}hang equation and universality class.
\newblock {\em Random Matrices: Theory and Applications}, 1:1130001, 2011.

\bibitem{PS2002.1}
M.~Pr\"ahofer and H.~Spohn.
\newblock Current fluctuations for the totally asymmetric simple exclusion process.
\newblock In V.~Sidoravicius, editor, {\em In and Out of Equilibrium:
  Probability with a Physics Flavor}, volume~51 of {\em Progress in
  Probability}, pages 185--204. Boston: Birkh\"auser, 2002.

\bibitem{Note1}
$f_{\protect\text{KPZ}}(y)$ is equal to $g''(y/2)/16$ in the notations of \cite{PS2002.1} and to $g''(y)/4$ in that of \cite{PrahoferWeb}.

\bibitem{vB2012.1}
H.~van Beijeren.
\newblock Exact results for anomalous transport in one-dimensional {H}amiltonian systems.
\newblock {\em Phys. Rev. Lett.}, 108:180601, 2012.

\bibitem{DL1998.1}
B.~Derrida and J.L. Lebowitz.
\newblock Exact large deviation function in the asymmetric exclusion process.
\newblock {\em Phys. Rev. Lett.}, 80:209--213, 1998.

\bibitem{D2007.1}
B.~Derrida.
\newblock Non-equilibrium steady states: fluctuations and large deviations of the density and of the current.
\newblock {\em J. Stat. Mech.}, 2007:P07023, 2007.

\bibitem{FS2006.1}
P.L. Ferrari and H.~Spohn.
\newblock Scaling limit for the space-time covariance of the stationary totally asymmetric simple exclusion process.
\newblock {\em Commun. Math. Phys.}, 265:1--44, 2006.

\bibitem{BR2000.1}
J.~Baik and E.M. Rains.
\newblock Limiting distributions for a polynuclear growth model with external sources.
\newblock {\em J. Stat. Phys.}, 100:523--541, 2000.

\bibitem{S2005.1}
T.~Sasamoto.
\newblock Spatial correlations of the 1{D} {KPZ} surface on a flat substrate.
\newblock {\em J. Phys. A: Math. Gen.}, 38:L549--L556, 2005.

\bibitem{J2000.1}
K.~Johansson.
\newblock Shape fluctuations and random matrices.
\newblock {\em Commun. Math. Phys.}, 209:437--476, 2000.

\bibitem{PrahoferWeb}
M.~Pr\"ahofer.
\newblock http://www-m5.ma.tum.de/{KPZ}.

\bibitem{BD2000.1}
E.~Brunet and B.~Derrida.
\newblock Probability distribution of the free energy of a directed polymer in a random medium.
\newblock {\em Phys. Rev. E}, 61:6789--6801, 2000.

\bibitem{PPH2003.1}
A.M. Povolotsky, V.B. Priezzhev, and Chin-Kun Hu.
\newblock The asymmetric avalanche process.
\newblock {\em J. Stat. Phys.}, 111:1149--1182, 2003.

\bibitem{GLMV2012.1}
M.~Gorissen, A.~Lazarescu, K.~Mallick, and C.~Vanderzande.
\newblock Exact current statistics of the asymmetric simple exclusion process with open boundaries.
\newblock {\em Phys. Rev. Lett.}, 109:170601, 2012.

\bibitem{B2009.1}
N.M. Bogoliubov.
\newblock Determinantal representation of the time-dependent stationary correlation function for the totally asymmetric simple exclusion model.
\newblock {\em SIGMA}, 5:052, 2009.

\bibitem{MSS2012.2}
K.~Motegi, K.~Sakai, and J.~Sato.
\newblock Long time asymptotics of the totally asymmetric simple exclusion process.
\newblock {\em J. Phys. A: Math. Theor.}, 45:465004, 2012.

\bibitem{P2014.1}
S.~Prolhac.
\newblock Spectrum of the totally asymmetric simple exclusion process on a periodic lattice - first excited states.
\newblock {\em J. Phys. A: Math. Theor.}, 47:375001, 2014.

\bibitem{P2015.2}
S.~Prolhac.
\newblock Asymptotics for the norm of {B}ethe eigenstates in the periodic totally asymmetric exclusion process.
\newblock {\em J. Stat. Phys.}, 160:926--964, 2015.

\bibitem{GS1992.1}
L.-H. Gwa and H.~Spohn.
\newblock Six-vertex model, roughened surfaces, and an asymmetric spin {H}amiltonian.
\newblock {\em Phys. Rev. Lett.}, 68:725--728, 1992.

\bibitem{P2015.3}
S.~Prolhac.
\newblock Current fluctuations and large deviations for periodic {TASEP} on the relaxation scale.
\newblock {\em J. Stat. Mech.}, 2015:P11028, 2015.

\end{thebibliography}

\begin{thebibliography}{10}

\bibitem{S_GS1992.1}
L.-H. Gwa and H.~Spohn.
\newblock Six-vertex model, roughened surfaces, and an asymmetric spin {H}amiltonian.
\newblock {\em Phys. Rev. Lett.}, 68:725--728, 1992.

\bibitem{S_P2014.1}
S.~Prolhac.
\newblock Spectrum of the totally asymmetric simple exclusion process on a periodic lattice - first excited states.
\newblock {\em J. Phys. A: Math. Theor.}, 47:375001, 2014.

\bibitem{S_MSS2012.2}
K.~Motegi, K.~Sakai, and J.~Sato.
\newblock Long time asymptotics of the totally asymmetric simple exclusion process.
\newblock {\em J. Phys. A: Math. Theor.}, 45:465004, 2012.

\bibitem{S_P2015.2}
S.~Prolhac.
\newblock Asymptotics for the norm of {B}ethe eigenstates in the periodic totally asymmetric exclusion process.
\newblock {\em J. Stat. Phys.}, 160:926--964, 2015.

\bibitem{S_P2015.3}
S.~Prolhac.
\newblock Current fluctuations and large deviations for periodic {TASEP} on the relaxation scale.
\newblock {\em J. Stat. Mech.}, 2015:P11028, 2015.

\bibitem{S_B2009.1}
N.M. Bogoliubov.
\newblock Determinantal representation of the time-dependent stationary correlation function for the totally asymmetric simple exclusion model.
\newblock {\em SIGMA}, 5:052, 2009.

\end{thebibliography}
\end{document}